\pdfoutput=1
\def\elsevier{elsevier}
\def\publisher{arxiv}

\ifx\publisher\elsevier
    \documentclass[5p]{elsarticle}
\else
    \documentclass{article}
\fi

\usepackage{graphicx}
\usepackage{amssymb}
\usepackage[utf8]{inputenc}
\usepackage{xcolor}
\usepackage{hyperref}
\usepackage{amsmath}
\usepackage[cmintegrals]{newtxmath}
\usepackage[official]{eurosym}
\usepackage{geometry}
\usepackage{array}
\usepackage{float}
\usepackage{verbatim}
\usepackage{framed}
\usepackage{multicol}
\usepackage{multirow}
\usepackage{ifthen}
\usepackage{caption}
\usepackage{footnote}
\makesavenoteenv{tabular}
\makesavenoteenv{table}

\if\publisher\elsevier

\else
    \usepackage{authblk}
    \usepackage[numbers,sort&compress]{natbib}
\fi

\newcolumntype{C}[1]{>{\centering\let\newline\\\arraybackslash\hspace{0pt}}m{#1}}
\graphicspath{{figs/}}
\ifx\publisher\elsevier
   \journal{Electrical Power and Energy System}
\fi

\begin{document}

\ifx\publisher\elsevier
    \begin{frontmatter}
\fi


%\title{B-FELSA: a tool to Benchmark Flexible Electric Loads Scheduling Algorithms considering the uncertainty in markets}
\title{Benchmarking Flexible Electric Loads Scheduling Algorithms under Market Price Uncertainty}




%\address{California, United States}

\ifx\publisher\elsevier
    \author[mymainaddress]{Koos~van~der~Linden}
    \author[mymainaddress]{Natalia~Romero}\author[mymainaddress]{Mathijs~M.~de Weerdt\corref{mycorrespondingauthor}}
    \cortext[mycorrespondingauthor]{m.m.deweerdt@tudelft.nl}
    
    \address[mymainaddress]{Faculty of Electrical Engineering, Mathematics and Computer Sciences, Delft University of Technology, Van Mourik Broekmanweg 6, 2628 XE Delft, The Netherlands}
\else
    \author{Koos~van~der~Linden}
    \author{Natalia~Romero}
    \author{Mathijs~M.~de Weerdt}
    \affil{Department of Software Technology, TU Delft}
    \maketitle
\fi    

\begin{abstract}
%Energy traders, policymakers and network operators are all interested in using the flexibility in energy consumption for sustaining stability in grids with intermittent and distributed generators.
Because of increasing amounts of intermittent and distributed generators in power systems, many demand response programs have been developed to schedule flexible energy consumption.
%In response, many algorithms for scheduling this flexibility have been developed.
%Power systems research has benefited from standards and defined case studies that enable benchmarking new methods.
However, proper benchmarks for comparing these methods are lacking, especially when decisions are made based on information which is repeatedly updated.
%The absence of these benchmarks creates a barrier for their implementation. 
This paper presents a new benchmarking tool designed to bridge this gap.
%Surveys that classify published offline and online flexibility planning algorithms, energy markets design, and demand response mechanisms were an input to define the benchmarking standards.
Surveys that classify flexibility planning algorithms were an input to define the benchmarking standards.
The framework can be used for different objectives and under diverse conditions faced by electricity energy stakeholders interested in flexibility scheduling algorithms.
It includes a simulation environment that captures the evolution of look--ahead information, which enables comparing online planning and scheduling algorithms.
The benchmarking tool is used to test seven planning algorithms measuring their performance under uncertain market conditions.
The results show the importance of online decision making, the influence of data quality on the algorithm performance, the benefit of using robust and stochastic programming approaches, and the necessity of trustworthy benchmarking.
\end{abstract}

\ifx\publisher\elsevier
\fi

\ifx\publisher\elsevier
    \end{frontmatter}
\fi

%%\linenumbers

\section{Introduction}
The integration of renewable energy is central for achieving energy security in a zero--carbon energy future~\cite{irea2018}. Some renewable energy sources are intermittent and difficult to forecast; thus, maximizing their use requires different operation and planning strategies to those traditionally used for more foreseeable or controllable loads and generators. Exploiting the flexibility in demand is a viable strategy for coping with the additional uncertainty~\cite{villar2018flexibility}. %Flexibility can be managed by offering incentives for users with flexible energy consumption, such as the financial advantage of paying for real--time energy market prices or the opportunity to bid for offering ancillary services. 

The scientific community has been responsive to this by developing a broad spectrum of algorithms to design effective incentive programs, commonly known as demand response (DR) programs. Over 70 publications of demand--side management were reviewed in~\cite{barbato2014optimization} to establish a general framework for such approaches; the authors analyzed whether users made selfish or cooperative decisions, the problem is solved with deterministic or stochastic methods, and the algorithms are offline versus online. Deng~et~al.\ summarized the objectives and issues in DR~programs~\cite{deng2015survey}, and Mukherjee and Gupta the control type considered in smart scheduling algorithms~\cite{mukherjee2015review}. 
%To assess the integration of different DR scenarios to the German market, Feuerriegel and Neuman published a comprehensive survey and framework of DR programs~\cite{feuerriegel2016integration}. 
One of the most extensive categorizations of DR~programs and algorithms was published in~\cite{vardakas2015survey}; it accounts for over 200~publications.

With so many options, power sector stakeholders need tools to compare and identify the method that suits their objectives the best. Unfortunately, only a small number of publications focuses on benchmarking existing methods or comparing new contributions to established ones. In~\cite{xu2016hierarchical}, a new approach that uses hierarchical control is compared to algorithms assuming central control, which are known for finding system--wide optimal solutions. Four models to maximize the value of flexible resources are proposed and compared in~\cite{kazempour2018value}. 

The challenge from assessing strategies increases when stakeholders consider real--time decisions, which require online algorithms, which update decisions based on new information.
%Different metrics for the assessment of online algorithms have been proposed and studied; for example, competitive ratio~\cite{sleator1985amortized,karp1992line}, Max/Max ratio~\cite{ben1994new}, and cooperative ratio~\cite{dorrigiv2008closing}. Such methods rely on definitions that are not shared by all problems making them difficult to use, in particular for algorithms applicable to real--life problems~\cite{dunke2017evaluating}, as flexible energy consumption scheduling.
%Offline algorithms assume that all information is available before executing them. Thus, the performance of these algorithms depends on their computing space and time, and the solution quality. In contrast, online methods are designed when information can be learned during execution, so their efficiency is determined by how well they use look--ahead information. This factor increases the complexity of measuring algorithms' efficiency. 
Simulation allows evaluating the performance of such online algorithms on multiple criteria.
%; thus, it is a suitable method to assess online algorithms for problems with many random variables and various objectives. 
Furthermore, simulation can help developing \emph{better} online optimization algorithms for complex dynamic problems~\cite{dunke2017evaluating}.
%''The study of the integration of distributed renewable generation, demand response, electric vehicles, or even aggregators in the electricity market [through simulation] is still very poor''~\cite{soares2018survey}. Dunke and Nickel and Soares~et~al. highlight the existence of a critical gap for the adoption of renewable integration solutions: a decision support tool to benchmark algorithms for flexible energy demand scheduling and to evaluate the policies associated with such algorithms~\cite{dunke2017evaluating,soares2018survey}.

This work presents a framework to benchmark offline and online algorithms for scheduling flexible energy loads. % and predict the outcome of (applicable) DR programs.
%and assess the effectiveness of market design for maximizing the use of flexible consumption, among other objectives.
This framework is the basis of an open-source toolbox, \emph{B-FELSA}. To compare results to the input data, check the reproducibility of observed behavior, and compare simulation outcomes using different algorithms, the toolbox includes a simulator. Users can change the input parameters, and use results to perform sensitivity analysis of their decisions. In summary, it is a verification environment, that offers tools for calibration and validation tasks making it an effective benchmarking tool~\cite{bialek2016benchmarking}.

This benchmarking framework is designed from the perspective of a consumer with flexible demand (or a stakeholder acting on his/her behalf) who wants to minimize total costs considering market incentives for load shifting. It covers shiftable and energy--based electric devices and accounts for the uncertainty associated with market incentives. 
%We assume the role of one of the stakeholders in the power sector, but it can be used by different stakeholders to simulate a specific situation and support their decisions. Such stakeholders include the consumer and aggregator of flexible loads, the distribution system operators (DSO), transmission system operators (TSO), and wholesale energy market policymakers. 
%%EXTRA SENTENCES
%%Methods, algorithms or solutions are not a complete tool for the applied sector. Stakeholders that would like to use such tools need standards to compare the performance of different methods and decide for the best.
%%The \emph{FlexEnergyBox} considers participation in wholesale markets, balancing and reserves markets. Participation in the first two corresponds to real--time pricing, and in the latter reserves markets- to demand bidding and buyback.
%%Such a prolific environment has been in response to the demand for solutions to cope with the increasing uncertainty in energy generation and supply.
This framework is used to compare algorithms that schedule this flexible energy demand. Straightforward and advanced heuristics, deterministic optimization and stochastic approaches are tested in a realistic scenario comparable to the Dutch energy market, as well as in an artificial scenario in which the quality of the provided data can be controlled.

The rest of this paper is organized as follows: Section 2 describes and motivates the scheduling context of the benchmark. Section 3 describes the framework and how it addresses the challenge of benchmarking online algorithms. This includes a description of the planning algorithms that are evaluated. Section 4 presents two case studies and discusses the results. Section 5 provides an overview of other energy decision support tools and explains the added value of the presented benchmarking framework. Section 6 concludes this research.

\section{Loads Scheduling Context}
Transmission system operators (TSOs) as well as distribution system operators (DSOs) around the world have identified the benefit from DR~programs to prevent congestion or ensure balance of supply and demand.
This has led to a large number of different programs, tariffs and market designs, which are employed to elicit the flexibility in loads from mainly residential, commercial customers, or both.
First we explain and motivate which DR context we have chosen for the benchmarking of loads scheduling algorithms, and then we give the type of flexible load that is represented.

%to exploit the flexibility of residential, commercial, as well as industrial loads
%In many countries, industries have various energy and network--use tariffs for direct DR~programs. Aggregators and retailers manage the flexibility  A retailer can schedule flexible consumption to self--balance its total energy demand and avoid paying the transmission system operator (TSO) for deviations from day--ahead (DA) commitments. Additionally, retailers or aggregators can trade flexibility in the electricity and reserves markets when a transaction is profitable. 

%Fig.~\ref{fig:Sthol} shows a representation of the interactions between power sector stakeholders concerning DR~programs. Such stakeholders include the user and aggregator of flexible loads, distribution system operators (DSO), TSOs, and wholesale energy market designers.

% \begin{figure*}[bt]
% \centering
% \includegraphics[width=\textwidth]{stakeholders.pdf}
% \caption{Relations between stakeholders and their interests. Adapted from \cite{feuerriegel2016integration}}
% \label{fig:Sthol}
% \end{figure*}

%Algorithms facilitate making decisions regarding the use of this flexibility to make the best of the economic reward that markets could offer. This section presents the different DR programs and market designs that have been considered when designing new algorithms.
%%Since the type of flexibility can affect the DR~program selection, we also introduce the different type of loads. %%The toolbox does not include all the different variants, so we highlight those that are covered by \emph{B-FELSA}.

\subsection{Demand response programs}
As described in the surveys on demand side management and demand response~\cite{barbato2014optimization,deng2015survey,vardakas2015survey}, DR~programs include centralized and decentralized control, advanced tariffs such as time of use, critical peak, real--time pricing and tier tariffs, to short--term electricity markets.

%When choosing the best DR~strategies, managers consider their objectives, the technology available and the type of flexible loads that they desire to influence. Peak clipping, valley filling, or load shifting objectives influence the design of DR~programs. The type of incentive offer to the consumer also helps differentiating DR~programs. 
%Incentive--based programs include direct control of the load, consumption curtailment or reduction and demand bidding and buyback~\cite{deng2015survey}. In these cases, the consumer receives compensation for demand change triggered by system contingencies. Demand bidding and buyback programs offer a reward for curtailment of a scheduled consumption at a specific bid price~\cite{deng2015survey}. This mechanism corresponds to offering ancillary services. 
%Time of use, critical peak pricing, real--time pricing and tier tariffs are well--known price--based programs.

%In this analysis we consider algorithms that schedule flexible loads based on real--time energy prices or the opportunity from offering reserves. Concatenating the demand scheduling with the market participation decision should improve the balancing effect of DR~programs. Hence, we consider DR~programs that reproduce the best market incentive.

%\subsection{Energy and reserve markets}

Considering current practice, the short--term electricity and ancillary services markets seem to be the preferred way of trading flexible demand. 
The most common of these markets consist of a two--sided auction where parties trade energy supply and demand in hourly or block intervals called program time units (PTUs)~\cite{brijs2017interactions}, and typically start about half a day before the day of delivery. Such an auction is known as the day--ahead (DA) market.
%After the DA market closure, the intersection of supply and demand curves for each hour determines the DA energy market hourly price.
%In this market balancing responsible parties commit to supplying or using the exact amount of energy that they traded per hour. 
%After aggregating the information from DA commitments, TSOs decide the balancing capacity needed for maintaining the grid's stability. 

Some countries have market designs that allow trading between stakeholders closer to the time of delivery. These markets are known as intraday markets in European countries. They facilitate trading continuously, or sometimes in discrete slots, after closure of the DA market. A similar environment is known as the adjustments market by the Electric Reliability Council of Texas~(ERCOT).

Parallel to DA and intraday markets, many TSOs use (voluntary) ancillary services to maintain grid stability. Trading parties can make bids to offer reserves supporting regulation up or down. Regulation up occurs in any instance when the demand exceeds the supply. The opposite case is known as regulation down. The deadline to place bids for offering ancillary services varies across countries, and conditions are typically different for offering voluntary or contracted services.
A bid typically consists of both a volume and a price, possibly different for each PTU.
%Voluntary offers are restricted to DA in some markets, but can be traded up to a few program time units (PTU) of delivery time, in other locations.
Offering ancillary services is equivalent to a demand bidding and buyback DR program, but with the option of also increasing consumption.

%After the time of delivery, responsible parties can adjust inconsistencies between actual supply or demand and their commitments. The adjustments are mostly paid to the TSO or may be executed through bi--lateral trading with other responsible parties experiencing the opposite imbalance. 
After energy delivery, differences between actual energy use and the DA commitment are settled by paying an imbalance price to the TSO. This is called the imbalance market.

In the benchmark we consider participation in the DA, imbalance and reserves markets. Intraday participation is not included but could be developed using the same logic used for online decisions in the imbalance market. 

\subsection{Type of load and its flexibility}
Electric devices in the grid can be classified in fixed, shiftable, and elastic energy--based or comfort--based~\cite{barbato2014optimization}. Shiftable devices have a fixed load profile that can only move in time. Elastic energy--based devices must consume a fixed amount of electricity within a given window; and comfort--based devices have flexible profiles only constrained by a comfort level target, e.g., temperature control systems~\cite{barbato2014optimization}. Furthermore, devices with energy storage capacity can offer unidirectional or bidirectional services (when devices can sell energy to the system)~\cite{mukherjee2015review}.

The analysis presented here is restricted to electric vehicle (EV) charging. EVs have relatively high flexibility and capacity, and this makes them good candidates for providing DR.
Moreover, they are representative of other elastic energy--based devices such as heat pumps.

%In addition to determining the best DR~program objective, stakeholders need a quantifiable benchmark, such as the total system utility, or household costs. Targets can be more specific to make them responsive to policies or customer demands. The amount of renewable energy or local generation use and the level of discomfort to flexible load users are examples of more complex targets~\cite{yao2017real}. Regardless of the objective, decision--makers need to consider the environment where the flexibility from DR~programs can be valorized. 

\section{Overview of \emph{B-FELSA}}
The previous section highlights the broadness of the scope of the DR program context. The large variety in the problem context has resulted in a large number of solution methods. This also makes comparing and benchmarking solution methods a hard task. \emph{B-FELSA} is designed to deal with this complexity.

\begin{figure*}[bt]
\centering
\includegraphics[width=\textwidth]{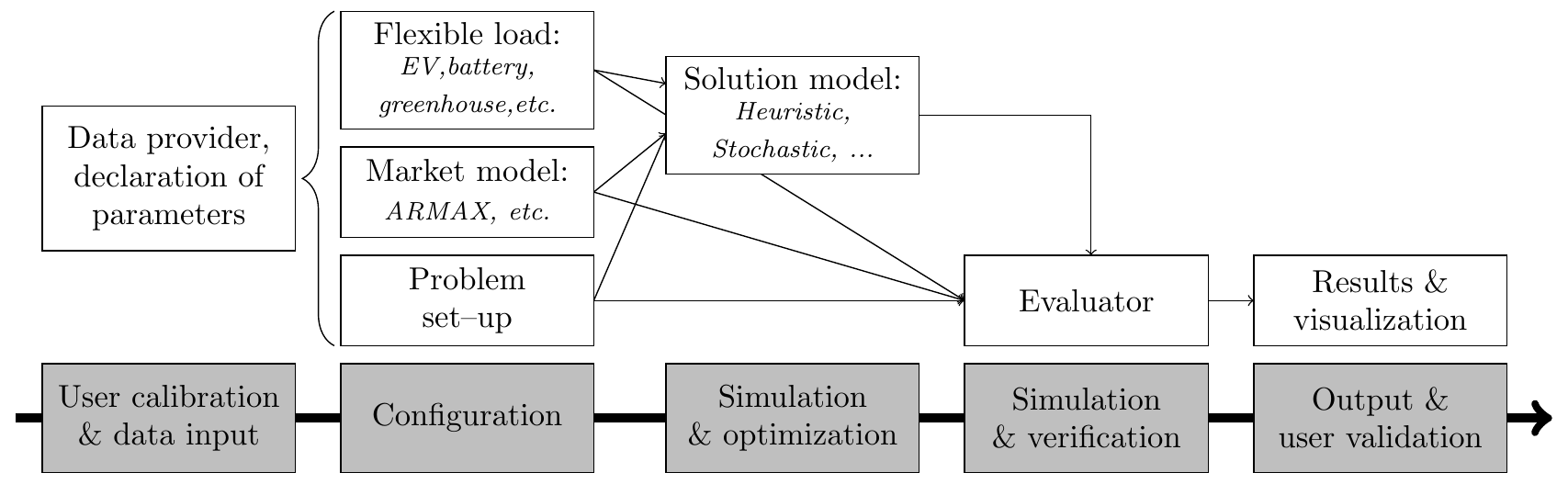}
\caption{\emph{B-FELSA} design}
\label{fig:design}
\end{figure*}

\emph{B-FELSA} has a modular design, which makes it a flexible platform for testing, comparing and designing new algorithms in various market design settings. The modules are grouped based on the steps followed for evaluating algorithms. The first phase, i.e.\ the input phase, is managed by the \emph{data provider}. In the following step, the experiments are configured using the input data and parameters. Three modules are responsible for this: \emph{flexible load}, \emph{market environment} and \emph{problem set--up}. The latter module refers to additional input parameters that define the solution model or external options, such as the existence of physical grid constraints. The next two phases and modules are critical for benchmarking algorithms. They use the market simulation environment to optimize scheduling decisions and measure the outcomes of those decisions. The modules are named \emph{solution model} and \emph{evaluator}, respectively. The last phase outputs results and offers a diverse set of \emph{visualizations} for results validation. Fig.~\ref{fig:design} illustrates the structure of \emph{B-FELSA}. \emph{B-FELSA} is open-source and its source can be obtained at \cite{bfelsa2019}.

Each of these modules is discussed in a separate section below.

\subsection{Data provider}
The input data consists of the loads, market, and model and simulator configuration. 
%The baseline load is an electric vehicle (EV) but the applications of the algorithms can be extended to other flexible loads with similar properties.
The market data includes DA market prices, regulation price and ancillary service usage. Additional data includes information such as grid constraints, and parameters defined by the user to model the uncertainty and its evolution.

\subsection{Flexible load}
The flexible load module is designed for EV charging, but with the option of extending it with other energy--based elastic or time--shiftable devices. The algorithms included in \emph{B-FELSA} consider a single type of flexible load that can be aggregated. The user can implement changes to \emph{B-FELSA} to account for a mixed pool of flexible loads as explained in~\cite{xu2016hierarchical}, or to consider micro--grid cost minimization or utility and maximization problems.

\subsection{Market data for simulator}
%Scheduling decisions are made considering the stakeholders' objectives, energy demand conditions, and electricity grid physical capacity. Additionally, stakeholders buy energy from different markets to maximize their surplus. 
%\emph{B-FELSA} assumes that 
Because DA market prices are relatively easy to predict, in the benchmark these are modeled deterministically, but 
%The intraday market is not considered but could be included following the same methodologies used for modeling real--time market prices. 
the uncertainty associated with the final regulation condition (i.e., whether ancillary services are accepted), and real--time energy prices are modeled through scenarios.
%These scenarios are generated for each combination of decision time and delivery time.
%According to the perspective adopted in this research for scheduling flexible energy consumption, 
%Decision parties need market energy prices as well as the amount of deployed ancillary services.

The simulation environment consists of a series of discrete trading times from the DA to delivery time. With each time step the information is updated. All decisions made for energy delivery in previous steps are fixed, but all other decisions can be updated. All decisions must be aggregated to verify that the total energy demand is fulfilled within the load constraints.

The user can generate scenarios to simulate the decision environment using different methods.
One of the models included to generate market price scenarios is the auto--regressive moving average model with exogenous variables (ARMAX), which is widely used for scenario generation and DA market price forecasting~\cite{conejo2010decision,olsson2008modeling,ansari2015coordinated,klaeboe2015benchmarking,alipour2017stochastic}. In particular, we follow and adapt the method used by Olsson and Soder~\cite{olsson2008modeling} for real--time balancing market prices generation. This method was identified as one of the best performing by Klaeboe~\cite{klaeboe2015benchmarking}. Both~\cite{olsson2008modeling,klaeboe2015benchmarking} focus on the Nordic power market prices. Olsson's method is adapted by using a Box-Cox-transformation~\cite{box1964analysis} instead of a log-transformation to normalize the data, and by using exogenous variables to capture the seasonal (daily) component~\cite{deen2019increasing}. 
%The toolbox includes parameters to create the simulation environment, but for a different data set, users should calibrate the parameters.
%Therefore, users must assess how well this method extends to their interest.

Ancillary services are used when the supply and demand need to be balanced. Acceptance happens on a continuous basis, in contrast to the discrete program time units (PTU) for offering the service. 
Furthermore, the required volume also determines acceptance.
Therefore, we use an abstraction that represents these two conditions. For each PTU, we estimate the percentage of time that the total energy offered is completely accepted and deployed, as is similarly done in~\cite{vagropoulos2013optimal}. Scenarios for this abstraction are also generated. We do this using a discrete Markov model with transition probabilities depending on the time of the day and season of the year. The transition probabilities are based on historic data. 

\begin{figure}[bt]
    \centering
    \if\publisher\elsevier
        \includegraphics[width=\columnwidth]{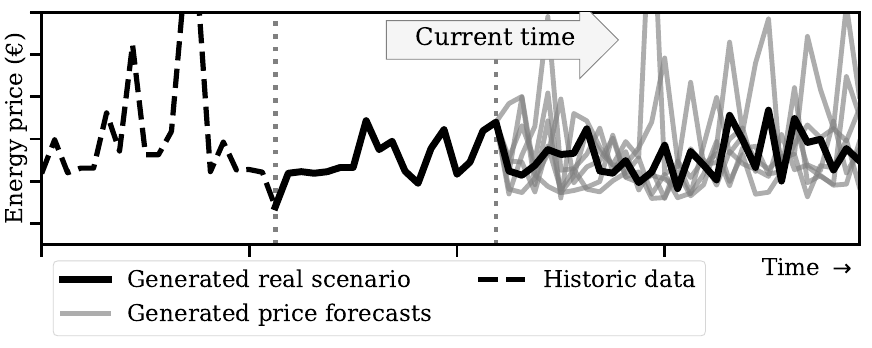}
    \else
        \includegraphics[width=0.7\textwidth]{generate_scenarios}
    \fi
    \caption{The ARMAX model is used to generate a real scenario (for evaluation) from historic data. The same ARMAX model is then used to generate a number of price forecasts from the generated scenario. These price forecasts are updated at every time step, starting from the current time (second vertical line in the figure).}
    \label{fig:scenario_generation}
\end{figure}

Fig.~\ref{fig:scenario_generation} shows how the ARMAX model is used to generate both an evaluation scenario (the real scenario) and price forecasts. During online evaluation, at every time step, these price forecasts are regenerated creating a series starting at the current time in the evaluation. Multiple scenarios are generated using Monte Carlo simulation. In the same way scenarios are also generated for the ancillary service usage using the Markov model introduced above. This models the market uncertainty and the increase in information as time draws closer to time of delivery.

\subsection{Solution model}
%Finding the energy consumption schedule that minimizes energy costs is a complex problem that has received extensive attention.
%These algorithms must be designed considering the application domain. 
An algorithm needs to decide the amount of energy purchased on the DA market per hour, and the energy use, the amount of up and down reserves committed, and the up and down reserve price bids per PTU. Every load has a session start and end, a starting state of charge (SOC), a required minimum SOC, a capacity, and a maximum energy usage (charging speed). The data provided to the solution models is the DA price per hour, and a set of scenarios consisting of an up and down regulating price, an imbalance price, and percentage of up and down reserve usages for every PTU. 

Solution methods include linear and mixed integer programming, non--linear programming, dynamic programming, particle swarm optimization, Markov process based--methods, evolutionary algorithms and other metaheuristics~\cite{mukherjee2015review,vardakas2015survey}. Additionally, game theory models can be used when considering solutions that rely on the stakeholders' interaction and their willingness to cooperate. 

\emph{B-FELSA} can evaluate all such algorithms except for multi--party cooperating and competition models. The following are already included in the toolbox:
\begin{enumerate}
\item The \emph{Direct} model (DI) represents buying energy immediately when plugged in.
\item The \emph{Optimal price} method (OP) minimizes the costs by charging at those times when the predicted energy market price is minimal. It does not provide reserves.
\item \emph{Heuristics} can be used to provide reasonable charging decisions. The MaxReg  heuristic method (MR) from \cite{sortomme2010optimal} is included. MR follows a preferred operating point (POP). With MR the POP is defined in such a way that it allows for maximum reserve participation. Reserve bids are quantity-only. For the analysis in this paper, the method has been updated to consider the reserve commitment deadline. It assumes that all its bids will be accepted and fully deployed and makes new robust bids based on this assumption.
\item With the \emph{Deterministic} model (DT), a user can plan energy and reserves market participation. The user determines a \emph{desired acceptance probability}, which is used to find the optimal quantity and price bid for bidding in the reserves market. It also optimizes DA, and imbalance markets participation. The algorithm is the interpretation by Van der Linden et~al.\cite{van2018optimal} of the solution method introduced by Sarker et~al.\cite{sarker_2016_optimal_0_235}.
\item The \emph{Quantity-only} method (QO) offers ancillary services but does not provide a price bid and assumes to be always accepted. It is implemented with DT by setting the desired acceptance probability to 100\%.
\item The \emph{Stochastic} optimization model uses a number of price scenarios to determine optimal reserve price bids and energy trade in DA and imbalance markets. Two versions are included:
\begin{enumerate}
\item A \emph{one stage} stochastic optimization model (SO1) that is similar to the deterministic model presented above, but optimizes for multiple scenarios instead of only the average scenario.
\item A \emph{two stage} stochastic optimization model (SO2) that uses binary variables to determine whether a price bid will be accepted or not. It was originally developed by Van der Linden et~al.\cite{van2018optimal} and improved by making the MIP formulation more tight and compact.
\end{enumerate}
\end{enumerate}
\emph{B-FELSA} also contains a perfect information (PI) solution model. Its solutions are obtained by providing SO2 with only the real scenario as input. The solution from PI can be used for comparison purposes.

\subsection{Evaluator}
%The combination of uncertainty in renewable energy sources and the daily decisions by generation managers, grid operators and other stakeholders determine the high complexity in modeling the grid's behavior in real--time. Reproducing this complexity in real cases is challenging. For reproducing and studying a system's behavior under normal conditions or hypothetical situations, computer simulation is known as the best method. Furthermore, discrete event simulation is also proposed as a holistic approach to evaluating online algorithms for dynamic problems~\cite{dunke2017evaluating}.

The evaluator uses simulation to measure the realisation of decisions made in every time step. Therefore, this module is critical to assess how online algorithms use new information to improve results.

The evaluator measures the run time and operation costs of the algorithm. It also measures the unmet demand and the exceeded battery capacity. It is possible that the algorithm makes a reserve commitment but the EV is not able to fulfill this commitment, because its battery capacity is reached. In this case the simulation continues as if it were possible, and the battery overflow is measured and outputted at the end. The same happens when the consumer's demand is not met.
%For battery capacity overflow the highest overflow during the complete charging period is measured (important in the case of V2G), and for unmet demand only the final charging level is measured.

\subsection{Visualization}
\emph{B-FELSA} provides different visualizations to the user with four objectives: first, comparing the market input data to the decisions made; second, the scheduling and bidding decisions as well as the consequences in terms of costs and risk; third, the evolution of decisions over time as well as the consequences of those decisions; and last, the ability to easily compare the effects of different algorithms, parameter settings or market settings side by side.

\section{Case studies}
The main purpose of \emph{B-FELSA} is to be able to compare different online solution methods for DR programs quantitatively. In this section \emph{B-FELSA} is used to study two case studies.
The case studies are used to answer the following questions:
\begin{enumerate}
    \item How does an online algorithm perform in comparison to the optimal decision under perfect information?
    \item What algorithm offers the best trade--off between scalability, and solution quality (feasible running time for minimum cost and risk of unmet demand)?
    \item What market participation strategy offers the minimum energy costs within my desired level of risk?
    \item How does the algorithm deal with new (better) information?
\end{enumerate}
%The case studies are performed from the perspective of the flexible energy consumer or aggregator, a price--taker who wants to minimize costs trading in energy and reserve markets, as well as to satisfy the load demand. This also means that in the analysis this consumer or aggregator is the sole decision maker of scheduling the flexibility in response to the market incentives.   

The first case study is a realistic case study with a market configuration similar to the Dutch energy market. It tests the performance of the algorithms mentioned in terms of operational costs, run time, and ability to deal with uncertainty.

The second case study is an artificial case study in which the (increase of the) prediction quality of the market realization can be controlled. It measures the effect of the (increase of) prediction quality (over time) on the algorithm's performance.

\subsection{Dutch energy market case study}
%The experimental setup for the Dutch case study is as follows. 
TenneT's market data~\cite{tennetpricedata} is used to create 95~historic scenarios. The ARMAX and Markov model are used to generate 10~scenarios per historic scenario, giving 950~of these scenarios in total. Then, at every increment of 15 minutes in every run, a total of 25~scenarios is generated using the same ARMAX and Markov model to represent forecasts. This (renewed) data is given to the algorithms to make or update its decisions. For each of these scenarios the overnight charging of one EV is simulated. The charging session takes 12 hours, so 48 time steps of 15 minutes. The EV has a battery capacity of 30kWh, an initial SOC of 1kWh, needs 26kWh, has a maximum charging speed of 7kW, and a charging efficiency of 90\%. Unmet demand is penalized with \euro{}60/MWh and battery capacity overflow with \euro{}200/MWh (in comparison, the average DA energy price is \euro{}32/MWh). Discharging, or vehicle-to-grid, is not allowed.

The Dutch energy market has DA prices per hour, and regulation and imbalance prices per PTU of 15~minutes. Here only voluntary secondary reserve bids are considered and these regulation bids are asymmetric and can be made up to 7~PTUs before delivery. Regulation is rewarded based on the amount of reserves deployed. Imbalance prices are based on the highest (lowest) price of the deployed reserve bids. The Dutch market has a minimum regulation bid size, but this is ignored here.

Based on experiments, the desired acceptance probability for DT is set to 50\%, and to 80\% for SO1. SO2 optimizes based on 20 scenarios.

All the solution methods were coded in Java. Gurobi 8.1.0~\cite{gurobi} is used as MIP solver for QO, DT, SO1 and SO2, with the MIP gap set to 1\%. Run time results are for an Intel~i7 6600U CPU with 8GB of RAM. 

\begin{figure*}[bt]
    \centering
    \includegraphics[width=\textwidth]{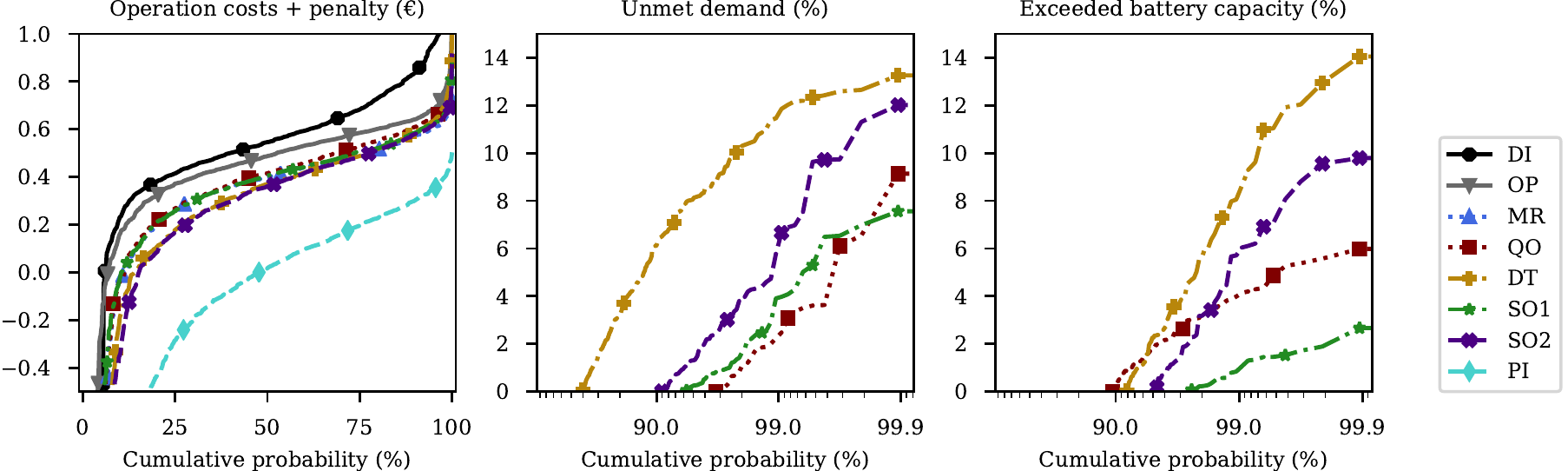}
    \caption{Results for the Dutch case study. The results show the performance for the $x$-th quantile in terms of operation costs plus penalty, unmet demand and exceeded battery capacity. Note that the horizontal axes of the second and third plot are logarithmic axes.}
    \label{fig:dutchcase}
\end{figure*}

\begin{table}[bt]
    \centering
    \caption{Results for the Dutch case study. The values shown are the mean $\pm$ the standard deviation of the results. The results are averages for one charging session.}
    \label{tab:dutchcase}
    \begin{tabular}{|l|c|c|c|c|} 
        \hline
            & Costs +           & Unmet             & Exceeded       & Run              \\ 
            & penalty           & demand            & capacity       & time             \\
            & (\euro{}) & (\%)  & (\%)              & (s)                              \\ \hline
        DI  & 0.47$\pm$0.51     & 0.0               & 0.0            & 1e--3$\pm$2e--3 \\ \hline
        OP  & 0.39$\pm$0.44      & 0.0              & 0.0            & 1e--3$\pm$1e--3   \\ \hline
        MR  & 0.27$\pm$0.46      & 0.0              & 0.0            & 1e-3$\pm$6e-3   \\ \hline
        QO  & 0.28$\pm$0.50      & 0.08$\pm$0.68    & 0.22$\pm$0.80  & 0.59$\pm$0.10   \\ \hline
        DT  & 0.21$\pm$0.54      & 1.63$\pm$2.84    & 0.33$\pm$1.51  & 0.58$\pm$0.08   \\ \hline
        SO1 & 0.27$\pm$0.48      & 0.11$\pm$0.69    & 0.02$\pm$0.21  & 0.66$\pm$0.10   \\ \hline
        SO2 & 0.19$\pm$0.58      & 0.24$\pm$1.14    & 0.17$\pm$1.03  & 73.8$\pm$41.2    \\ \hline
        PI  & -0.25$\pm$0.78     & 0.0              & 0.0            &                 \\ \hline
    \end{tabular}
\end{table}

Fig.~\ref{fig:dutchcase} and Table~\ref{tab:dutchcase} show the results of this case study. The values shown in the table are averages. Fig.~\ref{fig:dutchcase} is a quantile plot. A quantile plot shows the fraction of values (vertical axis) that fall below that quantile (horizontal axis). The unmet demand and exceeded battery capacity are reported as percentages of the requested load and battery capacity respectively. An average unmet demand of 1.6\% for example (the average unmet demand for DT) means that the battery is charged to 26.6kWh, 0.4kWh below its requested amount of 27kWh. With a charging speed of 7kW, this means that the EV missed approximately 4 minutes of charging time. This unmet demand is reflected in the cost with the penalization of \euro{}60/MWh, resulting in a penalty of 2.5 cents. 

The results for operation costs show only a small difference between the methods. A student's t-test proves that these differences are significant with $p<5\%$ except for MR, QO, and SO1. The differences in unmet demand are significant for all cases, except for again QO and SO1. Differences in exceeded battery capacity are also significant for all cases, except for QO and SO2. 
It can be concluded therefore that SO2 on average has the best results in terms of costs plus penalty. The difference between SO1 and DT shows that solving for multiple scenarios does not decrease operation costs, but does decrease unmet demand and exceeded battery capacity (this is also the case when SO1 is run with the desired acceptance probability set to 50\%). Interestingly, MR has cost results similar to QO and SO1 without having a penalty. What makes it more interesting is that MR does not use the scenario data at all. It also has a run time orders of magnitude smaller.

A general observation from the results is that there is a high variance in the results, and as a result, differences between methods are as small as 2\% of the total standard deviation. %0.01/0.52
The largest difference is between DI and SO2, with SO2 being 60\% cheaper than DI. %(0.47-0.19)/0.47
Having perfect information is on average yet another 85\% of the total standard deviation better in comparison to SO2.
But the standard deviation for PI is even higher than the total standard deviation. This means the performance variability is inherent to the problem. This also shows the importance of dealing with uncertainty in this problem.

\subsection{Controlled increasing prediction quality case study}
The setup of the case study with controlled increasing prediction quality is similar to the previous case study except for the following point. Instead of generating 25 scenarios per time step, the simulator generates $25q$ scenarios. From these $25q$ scenarios, 25 scenarios with the smallest error are selected. This error measure is defined in such a way that differences between the real and generated scenario at the beginning of a scenario have a higher weight.
%The weight function that is used here is $w(t) = 1 - t/T$ if $t < T$, and $w(t) = 0$  otherwise.
%This means that predictions for earlier time steps are more important, and predictions after time step $T$ are ignored in calculating the error. In this experiment, $T$ is set to six hours.
By changing the parameter $q$, the quality of the forecast can be regulated, with $q=1$ denoting no information increase, and higher $q$ means higher forecast quality.

\begin{figure*}[bt]
    \centering
    \includegraphics[width=\textwidth]{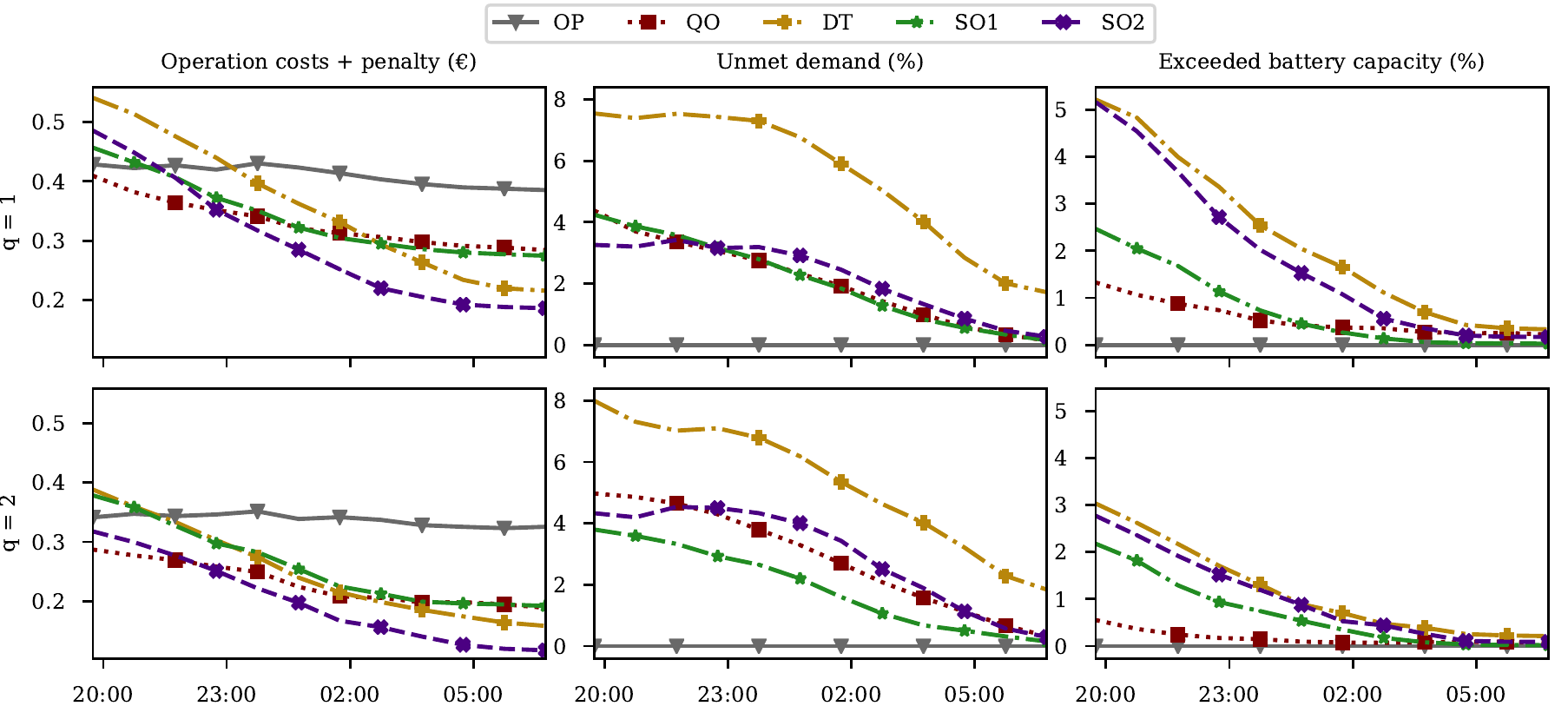}
    \caption{Results for the controlled increasing forecast quality case. Forecast quality is regulated by selecting the best 25 scenarios out of $25q$ generated scenarios. Over time the forecast quality increases. The figure shows the total end results per time step as if the decisions after this time step would not have been changed. In the case study, new decisions were made every 15 minutes. In the graph, the results are averaged per hour to increase readability.}
    \label{fig:increasecase}
\end{figure*}

Fig.~\ref{fig:increasecase} and Table~\ref{tab:inccase} show the effect of the increase in information over time. As time progresses, the methods are re-evaluated with more up to date data. The results in Fig.~\ref{fig:increasecase} shows what the end results would have been if no change in decisions was made from that point. MR, DI and PI have been left out from this evaluation.
MR is left out because the value of $q$ does not influence its decisions as MR ignores the data.
%The reasons for leaving out MR are twofold. First, MR does not use the provided data to make decisions, so its results for $q=2$ are the same as when $q=1$. Second, MR does not plan ahead, so it is not possible to show what its results would have been if its decisions had not changed.

A first glance shows directly that over time the operation costs for all methods decrease. The updated information allows the methods to improve their decisions and this gives better results. Fig.~\ref{fig:increasecase} shows that the four methods that provide reserves improve more over time than OP. This is because OP only optimizes based on the expected energy price. The other methods provide reserves and learn over time whether reserve bids are accepted and deployed or not, and can updated their decisions based on that afterwards.

In the previous case study, the heuristic MR method had similar performance than more complex methods like SO1, even though it does not use any of the provided data. But with improved data quality QO, DT, SO1 and SO2 all perform better on average in terms of penalized costs. This shows the importance of analyzing methods with different information quality.

The amount of exceeded battery capacity and unmet demand also decrease over time. The increased information quality makes only a small difference for the amount of unmet demand at the end of the charging session. For DT, SO1, and SO2 this difference is not statistically significant. Just as with $q=1$, QO, SO1, and SO2 all score well below 1\% unmet demand on average.
%1\% means only 2 minutes of missed charging time.
For these methods more than 90\% of the time there is no unmet demand or exceeded battery capacity at all. The difference in exceeded battery capacity is bigger and statistically significant for all methods expect for SO1, but SO1 already has almost no exceeded capacity when $q=1$.

\begin{table}[bt]
    \centering
    \caption{Results for the controlled increasing forecast quality case. The values shown are the averages for $q=2$ with the average difference between $q=2$ and $q=1$ in parentheses (see values in Table~\ref{tab:dutchcase}).}
    \label{tab:inccase}
    \begin{tabular}{|l|c|c|c|} 
        \hline
            & Costs +                       & Unmet                         & Exceeded                   \\ 
            & penalty (\euro{})             & demand (\%)                   & capacity (\%)              \\ \hline
        OP  & 0.33 (-0.06)  & 0.0           & 0.0            \\ \hline
%        MR  & 0.29          & 0.0           & 0.0            \\ \hline
        QO  & 0.19 (-0.09)  & 0.22 (+0.14)  & 0.06 (-0.15)   \\ \hline
        DT  & 0.16 (-0.06)  & 1.73 (+0.10)  & 0.21 (-0.12)   \\ \hline
        SO1 & 0.19 (-0.08)  & 0.11 (-0.00)  & 0.02 (-0.01)   \\ \hline
        SO2 & 0.12 (-0.07)  & 0.26 (+0.02)  & 0.09 (-0.08)   \\ \hline
    \end{tabular}
\end{table}

%With $q=2$ the cost decrease over time is higher than with $q=1$. {\color{red}Interestingly, QO even becomes the best method in terms of costs.} The difference between OP and the other methods that do provide reserves also becomes much smaller. This shows that the benefit of providing reserves becomes smaller when there is less uncertainty in imbalance price realization.

\section{Review of energy decision support tools}
A broad spectrum of tools and simulators have been developed to address the upcoming challenges for policy makers and other power sector stakeholders.
This section highlights the insights from four thorough reviews of computer tools developed in the last 20 years to inform decisions in the electric energy industry. Furthermore, it analyzes two tools that have capabilities comparable to \emph{B-FELSA}.

Already in 2009, Conolly et al.\ published a review of the different computer tools that supported decisions relevant to the integration of renewable energy~\cite{connolly2010review}. This survey aimed to inform decision--makers about the best tool to use for their specific situation. Tools were classified in: simulation, scenario, equilibrium--tool, top--down, bottom--up, operation optimization, and investment. None of these classifications focuses on assessing the strategies or algorithms used for scheduling loads.

Soares et al.\ classified simulation tools in electricity market, and micro--grid and smart grid simulators~\cite{soares2018survey}. Grid simulators are designed to support decisions on the physical integration of the power grid components, which is outside of the scope of \emph{B-FELSA}. 

Energy market simulators model the energy market as a series of interactions between different stakeholders negotiating electricity contracts. These simulators help those stakeholders to prescribe the consequences of their own decisions and those of other parties participating in the markets.  Hence, they could include the objectives of \emph{B-FELSA} as well. Some well--known examples are: the Simulator for Electric Power Industry Agents~(SEPIA)~\cite{harp2000sepia}, Powerweb, which includes both a market and grid model~\cite{zimmerman2004powerweb}, the Short-medium Run Electricity Market Simulator~(SREMS) based on game theory~\cite{migliavacca2007srems}, the  Electricity Markets Complex Adaptive System~(EMCAS)~\cite{north2003emcas}, the AMES, a wholesale power market test bed~\cite{li2009ames}, the advanced energy systems analysis model, EnergyPLAN~\cite{EnergyPLAN}, and the Multi-agent Simulator of Competitive Electricity Markets~(MASCEM)~\cite{vale2011mascem}. 
Many of these energy market simulators are very powerful and include agent models that can learn from their actions. Additionally, some have been enhanced by the integration with packages with important features. For example, ALBids is a decision support tool for agents negotiating in the energy market~\cite{pinto2012metalearning}. It includes different strategies already introduced in the literature. These strategies can be used by agents in MASCEM and compared considering different market environments.

Seventy--five modeling tools to analyze energy systems were studied by Ringkjøb et al.~\cite{ringkjob2018review}. Similar to~\cite{connolly2010review}, this more recent study highlights that energy computer tools are designed to answer specific questions. Ringkjøb et al.\ classify the tools in power systems analysis; operation decision support, which focus on dispatch problems, such as the unit commitment problem; investment decision support that assess full investment cycles; and scenarios for focusing on long--term industry projections~\cite{ringkjob2018review}. These classifications do not capture the aim of \emph{B-FELSA}. Nonetheless, among the tools included, the Distributed Energy Resources Customer Adoption Model Plus (DER-CAM+) has an objective relevant to our work~\cite{DERCAM+}. This tool classifies loads based on their fuel or end--use type. It integrates a power grid model with market data to represent energy, as well as reserve markets. Selecting the best strategy for distributed loads considering the market participation and grid constraints is one of its main objectives.

The tools studied in these four surveys have many applications, but they are mostly based on deterministic models~\cite{ringkjob2018review}. Furthermore, when agents can consider alternative strategies, these are predetermined or derived from the learning algorithms dependant on the market simulation~\cite{zhou2007agent}. Of the tools studied, ALBids and DER-CAM+ aim to find an optimal strategy for parties with distributed and flexible loads participating in energy markets. However, these tools include this capability among an array of options, which hinders their use for studying the effect from different factors (under controlled settings) on algorithms for scheduling decisions. 

\emph{B-FELSA} addresses the absence of a comparison methodology and simulation environment for algorithms that optimally schedule flexible loads. This is achieved by focusing on the decisions made by an agent trading energy to be used or dispatched by flexible loads. This agent is a price--taker. Thus, the uncertainty in the different markets is simulated through scenarios of prices and settlement decisions. Furthermore, we consider the implications of the strategic decisions made by flexibility traders for other important energy sector matters.

\section{Conclusions}
%%Flexible load scheduling involves significant uncertainty and from different sources. The uncertainty can be associated to renewable energy and its effect on energy prices, decisions by other stakeholders participating in energy markets, grid operation, and the flexible load constraints~\cite{soares2018survey}. The combination of DR programs objective, markets design, grid configurations, technology solutions, flexible load behavior, and user preferences creates\footnote{Needs a different verb} many solutions and high uncertainty.

Large numbers of DR programs and algorithms have been developed for dealing with increasing uncertainty in energy supply. This paper presents a benchmarking tool for quantitative analysis of these methods, called \emph{B-FELSA}. \emph{B-FELSA} is designed to be able to consistently compare flexible loads scheduling algorithms under a variety of different market designs, and varying information regarding future price and reserve acceptance scenarios.

The two presented case studies illustrate the value of \emph{B-FELSA}. The results show first that a simple analysis of algorithms based on expected values is not good enough, because of the highly volatile nature of imbalance prices and ancillary service usage.
%, performance results also show a high volatility. 
In particular, the strategy that uses expected values (DT) has the highest percentage of unmet demand and exceeded battery capacity.
%In order to arrive at statistically significant results, a large number of simulation runs is needed. By lack of enough historic data, this might mean that a good analysis requires a model to create these scenarios. 

Second, the case studies show that it is hard (very costly) to find schedules with reserves that are robust and meet the minimum energy and capacity requirements in all scenarios. This is of particular importance for stakeholders that want to minimize or quantify risk, or for a TSO that considers using flexible loads for maintaining system stability.

Third, \emph{B-FELSA} gives insight into how the solution methods improve their decisions over time. The uncertainty of the market behavior makes it difficult to trade based on schedules made day-ahead. The online analysis shows that updating your decisions during the day is necessary to decrease costs and to minimize risk, and that the quality of data should be considered when choosing a scheduling algorithm. It also shows the benefit of using stochastic programming or robust approaches, especially with regard to balancing risk and costs.

This paper focuses on the optimization from the perspective of flexible load consumers. With future work the presented framework should also be able to answer research questions from other stakeholders. The consumer is interested in minimizing costs and risk. Distribution system operators want to prevent congestion and maintain voltage level standards, and are therefore interested in consumer behavior and in providing the right incentives to consumers to maintain stability. Similarly, TSOs are interested in prescribing the balancing capacity available considering the incentives offered by the wholesale market and their imbalance settlements. And policymakers want markets that efficiently allocate resources and provide solutions that improve social welfare.
%This problem is a multi--objective and complex problem that may require multiple steps.

\emph{B-FELSA} can already answer some of the questions that these stakeholders might have. Still, there are a few important addition that would enlarge the scope of the proposed framework: the intraday market, other types of flexible load, and different types of local grid constraints are examples. The case studies also show that the (quality of the) data influences the performance of the algorithms. By adding other data generation methods it will be clearer what performance can be explained by the data, and what by the solution method. Other future work is to use the insight that \emph{B-FELSA} provides to make a better online scheduling algorithm.

Regulators must design the right incentive mechanism that enables social welfare of energy systems operations. Operators need to make the best term daily operations, maintenance, and planning decisions that guarantee a reliable and sustainable service. Users want the best and most affordable service. Moreover, users that can sell to the grid or can offer flexibility for a smooth operation may need an incentive and efficient planning algorithms to make their flexibility available. Therefore, there is a high demand for methods to support the diverse set of new decisions faced by the energy sector stakeholders. \emph{B-FELSA} provides the framework to assess these methods.

\begin{comment}
\begin{itemize}
\item Summarize work
\item Missing components: better market options; for example, intraday, other loads, market features
\item Future work: using lessons from simulator for better online method, assess DSO strategies to cope with local grid capacity constraints
\item Extend the the toolbox domain or integrate its methodology to major market simulators.
\end{itemize}

Regulators must design the right incentive mechanism that enables social welfare of energy systems operations. Operators need to make the best term daily operations, maintenance, and planning decisions that guarantee a reliable and sustainable service. Users want the best and most affordable service. Moreover, users that can sell to the grid or can offer flexibility for a smooth operation may need an incentive and efficient planning algorithms to make their flexibility available. Therefore, there is a high demand for methods to support the diverse set of new decisions faced by the energy sector stakeholders.

\end{comment}

\section*{Acknowledgments}
The work presented in this paper is funded by the Netherlands Organization for Scientific Research (NWO), as part of the Uncertainty Reduction in Smart Energy Systems program (URSES)



%% References


\ifx\publisher\elsevier
    \bibliographystyle{model1-num-names}
    \section*{References}
\else
    \bibliographystyle{unsrtnat}
\fi
\bibliography{sample.bib}

\begin{thebibliography}{38}
\providecommand{\natexlab}[1]{#1}
\providecommand{\url}[1]{\texttt{#1}}
\expandafter\ifx\csname urlstyle\endcsname\relax
  \providecommand{\doi}[1]{doi: #1}\else
  \providecommand{\doi}{doi: \begingroup \urlstyle{rm}\Url}\fi

\bibitem[{International Renewable Energy Agency}(2018)]{irea2018}
{International Renewable Energy Agency}.
\newblock Energy transition.
\newblock \url{https://www.irena.org/energytransition}, 2018.
\newblock Accessed: 2018-11-13.

\bibitem[Villar et~al.(2018)Villar, Bessa, and Matos]{villar2018flexibility}
Jos{\'e} Villar, Ricardo Bessa, and Manuel Matos.
\newblock Flexibility products and markets: Literature review.
\newblock \emph{Electric Power Systems Research}, 154:\penalty0 329--340, 2018.
\newblock \doi{10.1016/j.epsr.2017.09.005}.

\bibitem[Barbato and Capone(2014)]{barbato2014optimization}
Antimo Barbato and Antonio Capone.
\newblock Optimization models and methods for demand-side management of
  residential users: A survey.
\newblock \emph{Energies}, 7\penalty0 (9):\penalty0 5787--5824, 2014.
\newblock \doi{10.3390/en7095787}.

\bibitem[Deng et~al.(2015)Deng, Yang, Chow, and Chen]{deng2015survey}
Ruilong Deng, Zaiyue Yang, Mo-Yuen Chow, and Jiming Chen.
\newblock A survey on demand response in smart grids: Mathematical models and
  approaches.
\newblock \emph{IEEE Transactions on Industrial Informatics}, 11\penalty0
  (3):\penalty0 570--582, 2015.
\newblock \doi{10.1109/TII.2015.2414719}.

\bibitem[Mukherjee and Gupta(2015)]{mukherjee2015review}
Joy~Chandra Mukherjee and Arobinda Gupta.
\newblock A review of charge scheduling of electric vehicles in smart grid.
\newblock \emph{IEEE Systems Journal}, 9\penalty0 (4):\penalty0 1541--1553,
  2015.
\newblock \doi{10.1109/JSYST.2014.2356559}.

\bibitem[Vardakas et~al.(2015)Vardakas, Zorba, and
  Verikoukis]{vardakas2015survey}
John~S Vardakas, Nizar Zorba, and Christos~V Verikoukis.
\newblock A survey on demand response programs in smart grids: Pricing methods
  and optimization algorithms.
\newblock \emph{IEEE Communications Surveys \& Tutorials}, 17\penalty0
  (1):\penalty0 152--178, 2015.
\newblock \doi{10.1109/COMST.2014.2341586}.

\bibitem[Xu et~al.(2016)Xu, Callaway, Hu, and Song]{xu2016hierarchical}
Zhiwei Xu, Duncan~S Callaway, Zechun Hu, and Yonghua Song.
\newblock Hierarchical coordination of heterogeneous flexible loads.
\newblock \emph{IEEE Transactions on Power Systems}, 31\penalty0 (6):\penalty0
  4206--4216, 2016.
\newblock \doi{10.1109/TPWRS.2016.2516992}.

\bibitem[Kazempour and Hobbs(2018)]{kazempour2018value}
Jalal Kazempour and Benjamin~F Hobbs.
\newblock Value of flexible resources, virtual bidding, and self-scheduling in
  two-settlement electricity markets with wind generation—part i: Principles
  and competitive model.
\newblock \emph{IEEE Transactions on Power Systems}, 33\penalty0 (1):\penalty0
  749--759, 2018.
\newblock \doi{10.1109/TPWRS.2017.2699687}.

\bibitem[Dunke and Nickel(2017)]{dunke2017evaluating}
Fabian Dunke and Stefan Nickel.
\newblock Evaluating the quality of online optimization algorithms by discrete
  event simulation.
\newblock \emph{Central European Journal of Operations Research}, 25\penalty0
  (4):\penalty0 831--858, 2017.
\newblock \doi{10.1007/s10100-016-0455-6}.

\bibitem[Bialek et~al.(2016)Bialek, Ciapessoni, Cirio, Cotilla-Sanchez, Dent,
  Dobson, Henneaux, Hines, Jardim, Miller, et~al.]{bialek2016benchmarking}
Janusz Bialek, Emanuele Ciapessoni, Diego Cirio, Eduardo Cotilla-Sanchez, Chris
  Dent, Ian Dobson, Pierre Henneaux, Paul Hines, Jorge Jardim, Stephen Miller,
  et~al.
\newblock Benchmarking and validation of cascading failure analysis tools.
\newblock \emph{IEEE Transactions on Power Systems}, 31\penalty0 (6):\penalty0
  4887--4900, 2016.
\newblock \doi{10.1109/TPWRS.2016.2518660}.

\bibitem[Brijs et~al.(2017)Brijs, De~Jonghe, Hobbs, and
  Belmans]{brijs2017interactions}
Tom Brijs, Cedric De~Jonghe, Benjamin~F Hobbs, and Ronnie Belmans.
\newblock Interactions between the design of short-term electricity markets in
  the {CWE} region and power system flexibility.
\newblock \emph{Applied energy}, 195:\penalty0 36--51, 2017.
\newblock \doi{10.1016/j.apenergy.2017.03.026}.

\bibitem[van~der Linden(2019)]{bfelsa2019}
Koos van~der Linden.
\newblock Benchmark for flexible electric load scheduling algorithms v1.0.
\newblock \url{https://github.com/AlgTUDelft/B-FELSA}, 2019.

\bibitem[Conejo et~al.(2010)Conejo, Carri{\'o}n, Morales,
  et~al.]{conejo2010decision}
Antonio~J Conejo, Miguel Carri{\'o}n, Juan~M Morales, et~al.
\newblock \emph{Decision making under uncertainty in electricity markets},
  volume~1.
\newblock Springer, 2010.
\newblock \doi{10.1007/978--1--4419--7421--1}.

\bibitem[Olsson and Soder(2008)]{olsson2008modeling}
Magnus Olsson and Lennart Soder.
\newblock Modeling real-time balancing power market prices using combined
  {SARIMA} and markov processes.
\newblock \emph{IEEE Transactions on Power Systems}, 23\penalty0 (2):\penalty0
  443--450, 2008.
\newblock \doi{10.1109/TPWRS.2008.920046}.

\bibitem[Ansari et~al.(2015)Ansari, Al-Awami, Sortomme, and
  Abido]{ansari2015coordinated}
Muhammad Ansari, Ali~T Al-Awami, Eric Sortomme, and MA~Abido.
\newblock Coordinated bidding of ancillary services for vehicle-to-grid using
  fuzzy optimization.
\newblock \emph{IEEE Transactions on Smart Grid}, 6\penalty0 (1):\penalty0
  261--270, 2015.
\newblock \doi{10.1109/TSG.2014.2341625}.

\bibitem[Kl{\ae}boe et~al.(2015)Kl{\ae}boe, Eriksrud, and
  Fleten]{klaeboe2015benchmarking}
Gro Kl{\ae}boe, Anders~Lund Eriksrud, and Stein-Erik Fleten.
\newblock Benchmarking time series based forecasting models for electricity
  balancing market prices.
\newblock \emph{Energy Systems}, 6\penalty0 (1):\penalty0 43--61, 2015.
\newblock \doi{10.1007/s12667-013-0103-3}.

\bibitem[Alipour et~al.(2017)Alipour, Mohammadi-Ivatloo, Moradi-Dalvand, and
  Zare]{alipour2017stochastic}
Manijeh Alipour, Behnam Mohammadi-Ivatloo, Mohammad Moradi-Dalvand, and Kazem
  Zare.
\newblock Stochastic scheduling of aggregators of plug-in electric vehicles for
  participation in energy and ancillary service markets.
\newblock \emph{Energy}, 118:\penalty0 1168--1179, 2017.
\newblock \doi{10.1016/j.energy.2016.10.141}.

\bibitem[Box and Cox(1964)]{box1964analysis}
George~EP Box and David~R Cox.
\newblock An analysis of transformations.
\newblock \emph{Journal of the Royal Statistical Society: Series B
  (Methodological)}, 26\penalty0 (2):\penalty0 211--243, 1964.

\bibitem[Deen(2019)]{deen2019increasing}
Gerrit Deen.
\newblock Increasing the market value of wind power using improved stochastic
  process modeling and optimization.
\newblock Master's thesis, Delft University of Technology, 3 2019.

\bibitem[Vagropoulos and Bakirtzis(2013)]{vagropoulos2013optimal}
Stylianos~I Vagropoulos and Anastasios~G Bakirtzis.
\newblock Optimal bidding strategy for electric vehicle aggregators in
  electricity markets.
\newblock \emph{IEEE Transactions on power systems}, 28\penalty0 (4):\penalty0
  4031--4041, 2013.

\bibitem[Sortomme and El-Sharkawi(2010)]{sortomme2010optimal}
Eric Sortomme and Mohamed~A El-Sharkawi.
\newblock Optimal charging strategies for unidirectional vehicle-to-grid.
\newblock \emph{IEEE Transactions on Smart Grid}, 2\penalty0 (1):\penalty0
  131--138, 2010.

\bibitem[van~der Linden et~al.(2018)van~der Linden, de~Weerdt, and
  Morales-Espa{\~n}a]{van2018optimal}
Koos van~der Linden, Mathijs de~Weerdt, and Germ{\'a}n Morales-Espa{\~n}a.
\newblock Optimal non-zero price bids for evs in energy and reserves markets
  using stochastic optimization.
\newblock In \emph{Proceedings of the 15th International Conference on the
  European Energy Market (EEM)}, pages 574--578. IEEE, 2018.
\newblock \doi{10.1109/EEM.2018.8470023}.

\bibitem[Sarker et~al.(2016)Sarker, Dvorkin, and
  Ortega-Vazquez]{sarker_2016_optimal_0_235}
M.~R. Sarker, Y.~Dvorkin, and M.~A. Ortega-Vazquez.
\newblock Optimal {Participation} of an {Electric} {Vehicle} {Aggregator} in
  {Day}-{Ahead} {Energy} and {Reserve} {Markets}.
\newblock \emph{IEEE Transactions on Power Systems}, 31\penalty0 (5):\penalty0
  3506--3515, September 2016.
\newblock ISSN 0885--8950.
\newblock \doi{10.1109/TPWRS.2015.2496551}.

\bibitem[TenneT(2019)]{tennetpricedata}
TenneT.
\newblock Market information.
\newblock \url{http://www.tennet.org/bedrijfsvoering/ExporteerData.aspx}, 1
  2019.
\newblock Accessed: 2019-07-03.

\bibitem[{Gurobi Optimization Inc.}(2016)]{gurobi}
{Gurobi Optimization Inc.}
\newblock Gurobi optimizer reference manual.
\newblock \url{http://www.gurobi.com}, 2016.

\bibitem[Connolly et~al.(2010)Connolly, Lund, Mathiesen, and
  Leahy]{connolly2010review}
David Connolly, Henrik Lund, Brian~Vad Mathiesen, and Martin Leahy.
\newblock A review of computer tools for analysing the integration of renewable
  energy into various energy systems.
\newblock \emph{Applied energy}, 87\penalty0 (4):\penalty0 1059--1082, 2010.
\newblock \doi{10.1016/j.apenergy.2009.09.026}.

\bibitem[Soares et~al.(2018)Soares, Pinto, Lezama, and
  Morais]{soares2018survey}
Jo{\~a}o Soares, Tiago Pinto, Fernando Lezama, and Hugo Morais.
\newblock Survey on complex optimization and simulation for the new power
  systems paradigm.
\newblock \emph{Complexity}, 2018, 2018.
\newblock \doi{10.1155/2018/2340628}.

\bibitem[Harp et~al.(2000)Harp, Brignone, Wollenberg, and Samad]{harp2000sepia}
Steven~A Harp, Sergio Brignone, Bruce~F Wollenberg, and Tariq Samad.
\newblock {SEPIA}. a simulator for electric power industry agents.
\newblock \emph{IEEE Control Systems Magazine}, 20\penalty0 (4):\penalty0
  53--69, 2000.
\newblock \doi{10.1109/37.856179}.

\bibitem[Zimmerman and Thomas(2004)]{zimmerman2004powerweb}
Ray~D Zimmerman and Robert~J Thomas.
\newblock Powerweb: A tool for evaluating economic and reliability impacts of
  electric power market designs.
\newblock In \emph{IEEE PES Power Systems Conference and Exposition, 2004.},
  pages 1562--1567. IEEE, 2004.
\newblock \doi{10.1109/PSCE.2004.1397612}.

\bibitem[Migliavacca(2007)]{migliavacca2007srems}
G~Migliavacca.
\newblock {SREMS}: a short-medium run electricity market simulator based on
  game theory and incorporating network constraints.
\newblock In \emph{2007 IEEE Lausanne Power Tech}, pages 813--818. IEEE, 2007.
\newblock \doi{10.1109/PCT.2007.4538420}.

\bibitem[North et~al.(2003)North, Thimmapuram, Cirillo, Macal, Conzelmann,
  Boyd, Koritarov, and Veselka]{north2003emcas}
M~North, P~Thimmapuram, R~Cirillo, C~Macal, G~Conzelmann, G~Boyd, V~Koritarov,
  and T~Veselka.
\newblock {EMCAS}: An agent-based tool for modeling electricity markets.
\newblock In \emph{2003 Agent Conference on Challenges in Social Simulation},
  page 253, 2003.

\bibitem[Li and Tesfatsion(2009)]{li2009ames}
Hongyan Li and Leigh Tesfatsion.
\newblock The ames wholesale power market test bed: A computational laboratory
  for research, teaching, and training.
\newblock In \emph{2009 IEEE Power \& Energy Society General Meeting}, pages
  1--8. IEEE, 2009.
\newblock \doi{10.1109/PES.2009.5275969}.

\bibitem[Connolly et~al.(2013)Connolly, Lund, Mathiesen, {\O}stergaard,
  M{\"o}ller, Nielsen, Ridjan, Hvelplund, Sperling, Karn{\o}e,
  et~al.]{EnergyPLAN}
David Connolly, Henrik Lund, Brian~Vad Mathiesen, Poul~Alberg {\O}stergaard,
  Bernd M{\"o}ller, Steffen Nielsen, Iva Ridjan, Frede Hvelplund, Karl
  Sperling, Peter Karn{\o}e, et~al.
\newblock Smart energy systems: holistic and integrated energy systems for the
  era of 100\% renewable energy.
\newblock 2013.

\bibitem[Vale et~al.(2011)Vale, Pinto, Praca, and Morais]{vale2011mascem}
Zita Vale, Tiago Pinto, Isabel Praca, and Hugo Morais.
\newblock Mascem: electricity markets simulation with strategic agents.
\newblock \emph{IEEE Intelligent Systems}, 26\penalty0 (2):\penalty0 9--17,
  2011.
\newblock \doi{10.1109/MIS.2011.3}.

\bibitem[Pinto et~al.(2012)Pinto, Sousa, Vale, Pra{\c{c}}a, and
  Morais]{pinto2012metalearning}
Tiago Pinto, Tiago~M Sousa, Zita Vale, Isabel Pra{\c{c}}a, and Hugo Morais.
\newblock Metalearning in {ALBidS}: a strategic bidding system for electricity
  markets.
\newblock In \emph{Highlights on Practical Applications of Agents and
  Multi-Agent Systems}, pages 247--256. Springer, 2012.
\newblock \doi{10.1007/978-3-642-28762-6_30}.

\bibitem[Ringkj{\o}b et~al.(2018)Ringkj{\o}b, Haugan, and
  Solbrekke]{ringkjob2018review}
Hans-Kristian Ringkj{\o}b, Peter~M Haugan, and Ida~Marie Solbrekke.
\newblock A review of modelling tools for energy and electricity systems with
  large shares of variable renewables.
\newblock \emph{Renewable and Sustainable Energy Reviews}, 96:\penalty0
  440--459, 2018.
\newblock \doi{10.1016/j.rser.2018.08.002}.

\bibitem[{Berkeley Lab}(2016)]{DERCAM+}
{Berkeley Lab}.
\newblock Distributed energy resources customer adoption model plus
  ({DER-CAM+}) 2016-075.
\newblock \url{https://ipo.lbl.gov/lbnl2016-075/}, 2016.
\newblock Accessed: 2019-06-03.

\bibitem[Zhou et~al.(2007)Zhou, Chan, and Chow]{zhou2007agent}
Zhi Zhou, Wai Kin~Victor Chan, and Joe~H Chow.
\newblock Agent-based simulation of electricity markets: a survey of tools.
\newblock \emph{Artificial Intelligence Review}, 28\penalty0 (4):\penalty0
  305--342, 2007.
\newblock \doi{10.1007/s10462-009-9105-x}.

\end{thebibliography}






% \end{thebibliography}

\end{document}